# Interplay of Conventional and Spin-Exchange Auger Recombination in Magnetically Doped Quantum Dots

Valerio Pinchetti[1], Ho Jin [1,2], Clément Livache[1,3], and Victor I. Klimov[1]*

[1]Nanotechnology and Advanced Spectroscopy Team, C-PCS, Chemistry Division,
Los Alamos National Laboratory, Los Alamos, New Mexico 87545, United States

[2]Department of Chemistry, Ulsan National Institute of Science and Technology, Ulsan 44919, Republic of Korea

[3]Laboratoire de Physique de la Matière Condensée, Ecole Polytechnique, CNRS, IP Paris, 91128 Palaiseau, France

*Corresponding author's e-mail: klimov@lanl.gov

**Abstract**

Auger processes play a critical role in the behavior of colloidal quantum dots (QDs). While nonradiative Auger recombination is a key performance-limiting factor in light-emitting applications, Auger effects also present exciting opportunities for hot-carrier manipulation in emerging optoelectronic and photochemical technologies. Recent studies have shown that incorporating magnetic manganese (Mn) dopants into QDs enables an ultrafast spin-exchange (SE) mechanism that significantly accelerates Auger recombination and related processes. However, the fundamental dynamics of SE-Auger recombination – including its dependence on QD size, exciton multiplicity, and Mn-ion content – remain largely unexplored. Here, we investigate SE-Auger recombination in Mn-doped QDs across three regimes defined by the energetic alignment between the QD band-edge exciton ($E_X$) and the internal Mn spin-flip transition ($E_{Mn}$): *resonant* ($E_X \approx E_{Mn}$), *downhill* ($E_X > E_{Mn}$), and *uphill* ($E_X < E_{Mn}$). We find that SE-Auger recombination dominates multiexciton dynamics in both resonant and downhill regimes, proceeding on ultrafast, sub-picosecond timescales – significantly faster than conventional Auger decay. Notably, SE-Auger lifetimes are primarily governed by the occupancy of intrinsic excitonic states, with minimal dependence on QD size or the number of excited Mn ions. In the uphill regime, SE-Auger recombination coexists with conventional Auger processes, enabling direct comparison of their timescales alongside co-occurring hot-carrier cooling. These findings establish that SE-Auger recombination outpaces phonon-assisted cooling, making it uniquely suited for the generation and manipulation of hot carriers, and opening new avenues in advanced photoconversion and photochemistry.

**Introduction**

Auger recombination is a nonradiative process in which an electron recombines with a hole by transferring energy to a third (acceptor) charge carrier[1]. In bulk semiconductors, due to conservation of translational momentum, Auger decay exhibits thermally activated behavior with an activation energy proportional to the bandgap of the material[2]. Because of this, the rate of Auger decay decreases rapidly with decreasing temperature or increasing bandgap. In nanoscale semiconductor colloidal crystals or colloidal quantum dots (QDs), translation motion of charge carriers is suppressed leading to relaxation of the translation momentum conservation and the elimination of the activation barrier[2,3]. As a result, Auger decay rates are greatly increased. Moreover, they become insensitive to the semiconductor bandgap or other details of the electronic structure and exhibit a surprisingly universal material-independent scaling with the QD volume[4,5] ($V_{QD}$). In the case of a two electron-hole pair (biexciton) state, this scaling can be expressed as $\tau_{A,XX} = \chi V_{QD}$, where $\tau_{A,XX}$ is the biexciton Auger lifetime, and $\chi$ is a proportionality constant that is ~1 ps nm$^{-3}$ for most QDs studied including those made of direct- and indirect-gap semiconductors[6]. Based on this expression, $\tau_{A,XX}$ varies from a few picoseconds for small QDs (~1 nm radius) to ~300 ps for large particles (~4 nm radius).

Since Auger lifetimes are much shorter than the radiative time constants (typically, a few to tens of nanoseconds in II-VI and III-V QDs), Auger recombination is detrimental to applications requiring high emissivity of QDs, including light-emitting diodes[7,8] (LEDs) and especially lasers[6,9], since in later devices biexcitons are typically the primary optical-gain species. To impede Auger decay in QDs and thus reduce its negative impact on light-emitting devices, strategies such as controlling Auger interactions through proper heterostructuring[10,11] and introducing compositional gradients[12] have been developed.

On the other hand, there are promising applications of QDs that could benefit from high Auger recombination rates. Auger decay can be viewed as an *up-conversion* process in which an electron (or hole) is excited to a higher energy state by the transfer of energy from a recombining electron-hole pair. This process could be useful, for example, in photochemistry assisted by energetic *hot* electrons[13,14]. It can also be applied in photoemission (PE) if an electron excited by Auger energy transfer is ejected from a QD[15]. However, this requires at least two Auger excitation steps, since the electron affinity of a material ($E_a$) is usually higher than its bandgap (which determines the energy released during electron-hole recombination). Because of this, the Auger contribution to PE is usually weak[15]. Indeed, although enhanced by quantum confinement, the rate of Auger energy transfer is still slower than the rate of energy loss due to phonon emission[16]. Therefore, a hot electron created by the first Auger excitation step loses its energy via phonon emission before the second Auger energy transfer step occurs.

Recent studies of manganese (Mn) doped QDs have shown that the rate of Auger up-conversion can be dramatically enhanced by employing not direct but rather *exchange* Coulomb interactions mediated by Mn impurities[16-19]. In particular, due to the extremely short (sub-picosecond) time scales of energy exchange between the Mn-based and QD intrinsic states, it was possible to realize highly efficient PE via successive energy transfers from two excited Mn ions[20]. Since the Mn-QD energy exchange leading to electron emission was accompanied by transfer of 1 unit of spin (forward and backward) between the two species, the realized PE was termed spin-exchange (SE) PE.

Here we investigate the SE-Auger recombination process underlying SE-PE, focusing on how it is influenced by the relative positions the QD size-dependent band-edge exciton energy ($E_X$) and the internal spin-flip transition energy of a Mn ion ($E_{Mn}$). Specifically, we consider three distinct regimes: *resonant* ($E_X \approx E_{Mn}$), *downhill* ($E_X > E_{Mn}$), and *uphill* ($E_X < E_{Mn}$). The first two are dominated by sub-picosecond SE-Auger recombination, which is activated when a QD is excited with two-electron hole pairs. In the uphill regime, SE-Auger recombination coexists with conventional Auger decay: the former dominates the early time dynamics associated with higher-order multiexcitons (three or more), while the latter governs the later-time decay of biexcitons composed of band-edge carriers. Interestingly, the SE-Auger decay rate shows no apparent dependence on QD-size or the number of excited Mn ion. Instead, it is primarily determined by the occupancy of intrinsic QD excitonic states, which is controlled by their energetic alignment relative to the internal Mn spin-flip transition.

**QD Samples**

For these studies, we synthesized four Mn-doped CdSe/CdS QD samples with mean CdSe core radii ($r$) ranging from 1.25–2.5 nm, along with four similarly sized reference undoped samples. In all cases, the thickness of a protective CdS shell was approximately one semiconductor monolayer (see Fig. 1a for a schematical depiction of the QDs and Fig. 1b for their transmission electron microscopy images).

For both doped and undoped samples, photoluminescence (PL) is dominated by radiative recombination of intrinsic electron-hole pairs (excitons) (Fig. 1c, left). The corresponding exciton energy ($E_X$) can be inferred from the peak of the PL band, and for the QD sizes studied here, it varies from below ($r = 2.5$ nm) to above ($r = 1.25$ nm) the energy of the Mn spin-flip transition of $E_{Mn} = 2.1$ eV (Fig. 1c, left). As shown previously[16], the energy difference $\Delta_{X\text{-}Mn} = E_X - E_{Mn}$ governs the rate of energy exchange between intrinsic exciton and Mn-based states, and their relative occupancies.

Optical absorption spectra of undoped samples show a sharp band-edge 1S peak, which shifts to higher energies with decreasing QD size (Fig. 1c, right), following the same trend observed for the PL band. Upon Mn incorporation, the 1S peak and other absorption features broaden, a typical outcome of the diffusion doping process and generally attributed to increased size polydispersity[21,22].

To prepare Mn-doped samples, we used a diffusion doping procedure introduced in refs [21,23] and later refined in ref. [22] (see Methods for synthesis details). In this method, Mn is initially incorporated into the QD surface layer as MnSe, after which Mn ions diffuse into the QD interior, substituting internal Cd ions. According to elemental analysis using inductively coupled plasma optical emission spectroscopy (ICP-OES), the Mn content ranges from 15–30% (ref. [22]). This analytical technique detects both internally and surface-incorporated Mn ions. However, the latter exhibits only weak exchange coupling to QD electronic states, since the wavefunctions of the QD band-edge states decay rapidly toward the QD periphery.

To determine the content of internal Mn ions, which are strongly coupled to the intrinsic states of the QDs, we perform magnetic circular dichroism (MCD) measurements (see Methods and Supplementary Note 1 for details). This technique probes the magnetic-field-induced difference in transmission between left- and right-circularly polarized light. Based on the MCD data (Fig. 1d,e, and Supplementary Figs. 1–4), the Mn content in the synthesized samples ranges from 1.7% to 2.9%. As expected, these values are lower than those obtained from ICP-OES measurements. Importantly, however, even for the lowest Mn content estimated via MCD, the number of internal Mn ions is more than one – sufficient to produce a pronounced effect of Mn-QD exchange interactions on carrier dynamics[16].

The structural, spectroscopic, and magnetic characteristics of all undoped and Mn-doped QD samples are summarized in Supplementary Tables 1 and 2.

**SE-Driven Exciton Transfer**

In addition to influencing the optical properties of QDs through the effective magnetic field generated by polarized spins of magnetic impurities, Mn ions also participate in fast bidirectional energy (or excitation) transfer with intrinsic QD states[16]. In the ground state, the spins of all five 3d electrons in the Mn ion are co-aligned, resulting in a total ion's spin of $S_{Mn} = 5/2$. This corresponds to the so-called $^6A_1$ configuration, referred to here as the *spin-up* Mn state. In the excited Mn state, one of the 3D electron spins is flipped, reducing the total spin to $S_{Mn} = 3/2$. This state corresponds to the $^4T_1$ configuration, referred to here to as the *spin-down* Mn state.

Due to spin-selection rues, the spin-flip transition between the $^4T_1$ and $^6A_1$ states cannot be initiated by light[24]. However, it can occur efficiently via spin- and charge-conserving carrier exchange between the Mn ion and the QD intrinsic states[25]. When the energy is transferred to the Mn ion – initiating the $^6A_1 \rightarrow ^4T_1$ transition – we refer to this process as *forward* transfer. Conversely, the reverse process, in which energy is transferred from the Mn ion back to the QD (initiating the $^4T_1 \rightarrow ^6A_1$ transition), is referred to as *backward* transfer.

In Fig. 2a (left), we schematically illustrate the spin-conserving forward and backward transfers. During forward transfer, the energy of a QD exciton (X) is transferred to the Mn ion, exciting it to the $^4T_1$ state (denoted as Mn*). This process can be viewed as a sequence of two corelated spin transfers: a spin-up electron moves from the Mn ion's 3d shell to fill the valence-band (VB) hole state of the QD; this is immediately followed by a spin-down electron transfer from the QD conduction band (CB) into the Mn 3d shell. As a result, the Mn-ion undergoes a spin change of $\Delta S_{Mn} = 1$. Consequently, to conserve total spin, the QD exciton involved in the transfer must also have a spin of $S_X = 1$, which corresponds to the optically inactive, or *dark*, exciton state.

Assuming spin-1/2 for both the electron and hole, the exciton consists of four spin configurations – two with $S_X = 0$ (bright states) and two with $S_X = 1$ (dark states) (Fig. 2b). Due to this complex spin structure, there is always at least one excitonic spin component with the appropriate spin magnitude and orientation for efficient energy transfer to a Mn ion (Fig. 2b). Combined with the intrinsically strong Mn-CdSe exchange coupling[24,25], this enables extremely fast QD-Mn energy transfer occurring on a time scale of ~100 fs (ref. [16]).

The reverse process, or backward transfer, also begins with a Mn ion from which a spin-down electron moves into the QD CB, followed by a spin-up electron transfer from the QD VB to fill the vacancy ('hole') in the 3d Mn shell. A simplified excitonic representation of spin-conserving QD-Mn energy exchange is shown in Fig. 2a (right).

**SE-Auger Recombination versus Conventional Auger Recombination**

As recently demonstrated in ref. [20], a similar Mn–QD energy exchange process mediated by SE interactions enables ultrafast recombination of a hybrid biexciton composed of an intrinsic QD exciton and an excited Mn ion. This process – referred to as *SE-Auger recombination* – is schematically illustrated in Fig. 2c (left, SE representation) and Fig. 2c (right, excitonic representation). Similar to SE-mediated single-exciton transfer (Fig. 2a), the process involves two spin transfers: first, a spin-down electron from the 3d shell of the excited Mn ion moves into a high-energy excited state of the QD CB; this is followed by a spin-up electron transfer from the QD band-edge CB state into the Mn 3d shell. As a result, the Mn ion returns to its ground-state spin-up configuration, while the band-edge QD exciton is promoted to a highly excited, or *hot*, state.

We begin our analysis of SE-Auger recombination with a QD sample of radius $r$ = 1.75 nm, for which the band-edge excitonic transition $E_X$ = 2.1 eV is resonant with the internal Mn spin-flip transition (see optical spectra in Fig. 1c, second dataset from the bottom). To monitor QD population dynamics, we use a transient absorption (TA) pump-probe technique, in which changes in the optical absorption coefficient ($\Delta\alpha$) induced by a short 110-fs pump pulse – resolved in both time and spectral domains – are tracked as a function of time using a variably delayed broad-band probe pulse of a femtosecond while-light continuum.

For all measurements, the pump photon energy ($h\nu_p$) was 2.41 eV and the per-pulse pump fluence ($w_p$) ranged from 0.001 to 2.2 mJ cm$^{-2}$. The QD samples were prepared as hexane solutions, loaded into 1-mm pathlength optical cells, and vigorously stirred throughout the measurements to prevent sample degradation and eliminate TA artifacts caused by uncontrolled photocharging[26,27]. All TA measurements were conducted at room temperature (see Methods for experimental details).

In Fig. 3a, we present the pump-fluence-dependent TA dynamics of a Mn-doped 'resonant' sample, recorded at the position of the lowest-energy 1S TA peak (Supplementary Fig. 5). This feature is primarily attributed to state-filling-induced bleaching of the $1S_e$ electron state[28]. Therefore, the dynamics of this signal ($\Delta\alpha_{1S}$) can be used to monitor the average occupancy of the band-edge exciton state ($\langle N_X \rangle$).

For the data shown in Fig. 3a, $w_p$ varied from 0.9 to 890 μJ cm$^{-2}$, corresponding to an average per-dot excitonic occupancy of $\langle N \rangle$ = 0.021 to 21. This was calculated using the relation $\langle N \rangle$ = ($w_p\sigma_{abs}/h\nu_p$), where $\sigma_{abs}$ = 9.1×10$^{-15}$ cm$^{-2}$ is a QD absorption cross-section at $h\nu_p$ = 2.41 eV (see Methods for the experimental determination of $\sigma_{abs}$). At low, pump levels, corresponding to sub-single-exciton QD occupancies ($\langle N \rangle \ll 1$), the TA traces show no noticeable decay over the 12-ps time window shown in Fig. 3a. However, as ⟨N⟩ approaches and then exceeds 1, the TA dynamics develop a fast initial sub-picosecond decay component, which emerges at ⟨N⟩ of ~1, as indicated by the pump-fluence dependence of the ratio of the early time TA amplitude (*A*) and the slow single exciton background (*B*) (Fig. 3a, main panel and inset). This suggests that the appearance of the fast signal correlates with onset of the regime where more than one exciton is injected into the QD.

The sub-picosecond depopulation dynamics is a hallmark of SE-Auger recombination of a hybrid multiexciton state that includes both intrinsic QD and Mn-based excitations. This phenomenon was previously reported in ref. [20] and is illustrated by the mechanism shown in Fig. 2c and the bottom inset in Fig. 3a. To determine the SE-Auger decay time constant ($\tau_{SE-A}$), we isolate multiexciton dynamics by subtracting the slow single-exciton background from the time transients measured at ⟨N⟩ > 1 (Fig. 3b). The resulting traces are then fitted to a single-exponential decay convoluted with a Gaussian function representing the 150-fs-wide instrument response function of our TA setup. Using this approach, $\tau_{SE-A}$ is found to be ~1 ps for ⟨N⟩ values between 2.0 to 3.7, and it gradually decreases to 230 fs at ⟨N⟩ = 21 (Fig. 3b, inset).

It is instructive to compare SE-Auger dynamics in Mn-doped QDs with the conventional Auger dynamics in an undoped reference sample. In the latter case (Fig. 3c and Supplementary Fig. 6), Auger recombination also appears as a fast component in the TA decay, emerging at ⟨N⟩ > 1 (Fig. 3c, inset). However, its time constant ($\tau_A$) is approximately 90 ps – more than two orders of magnitude longer than $\tau_{SE-A}$. This stark difference highlights the exceptionally strong exchange coupling between magnetic impurities and the semiconductor host in Mn-doped CdSe QDs[25], leading to a dramatic acceleration of Auger dynamics due to the involvement of SE interactions.

**Assessment of Relative Occupancies of Intrinsic QD and Mn-Based States**

To rationalize the observed scaling of $\tau_{\text{SE-A}}$ with $\langle N \rangle$, we analyze the distribution of the initially injected excitons between the intrinsic QD states and Mn-based states, with their average populations denoted as $\langle N_{\text{X}} \rangle$ and $\langle N_{\text{Mn}} \rangle$, respectively. For this purpose, we compare the early-time 1S bleach amplitudes as a function of $\langle N \rangle$ for both the Mn-doped and undoped reference QD samples (Fig. 3d). We express $\Delta\alpha_{1S}$ as a normalized quantity, $q_{1S} = |\Delta\alpha_{1S}|/\alpha_{0,1S}$, where $\alpha_{0,1S}$ is the ground-state absorption coefficient of the 1S transition. The value of $q_{1S}$ is directly related to the average occupancy of the $1S_e$ electron state, $\langle N_{e,1S} \rangle$, by the expression $q_{1S} = \langle N_{e,1S} \rangle/2$ (ref. [28]). Because the $1S_e$ level is twofold degenerate, its maximum occupancy is 2, yielding a maximum $q_{1S}$ of 1.

To relate $q_{1S}$ to the average number of photoinjected excitons per dot ($\langle N \rangle$), we assume that the QD excitonic occupancy follows Poisson statistics, which is typical for short-pulse excitation above the band-edge transition[28]. In this case, the probabilities of a QD being occupied by exactly one exciton ($p_1$) or two or more excitons ($p_{\geq 2}$) are given by: $p_1 = \langle N \rangle e^{-\langle N \rangle}$ and $p_{\geq 2} = 1 - p_0 - p_1 = 1 - e^{-\langle N \rangle}(1 + \langle N \rangle)$. Taking into account that $\langle N_{e,1S} \rangle = p_1 + 2p_{\geq 2}$, we derive the following relationship between $q_{1S}$ and $\langle N \rangle$: $q_{1S} = 1 + e^{-\langle N \rangle}(\langle N \rangle/2 - 1)$.

As shown in Fig. 3d, the experimental dependence of $q_{1S}$ on $\langle N \rangle$ measured for the undoped QDs (open squares) closely follows the Poissonian predication (solid line). In contrast, the $q_{1S}$ values measured for the Mn-doped sample (Fig. 3d, solid squares) lie systematically below both the theoretical curve and the data from the reference sample. This deviation indicates a partial loss of population from QD intrinsic states due to excitation transfer to Mn ions.

To quantify the distribution of photoinjected excitons between intrinsic and Mn-based states, we use the experimental $q_{1S}$-versus-$\langle N \rangle$ dependence from the reference sample as a calibration curve to relate 1S bleach magnitude to the actual number of excitations remaining in the intrinsic QD states, $\langle N_X \rangle$. We then determine $\langle N_X \rangle$ for the doped sample using the condition: $q_{1S,\text{ref}}(\langle N_X \rangle) = q_{1S,\text{Mn}}(\langle N \rangle)$, where $q_{1S,\text{ref}}$ and $q_{1S,\text{Mn}}$ are the $q_{1S}$-versus-$\langle N \rangle$ dependences measured for the reference and Mn-doped samples, respectively. This procedure is illustrated graphically in Fig. 3d via three projections: $\langle N \rangle \rightarrow q_{1S,\text{Mn}} \rightarrow q_{1S,\text{ref}} \rightarrow \langle N_X \rangle$.

Once $\langle N_X \rangle$ is determined, the number of excited Mn ions is calculated as $\langle N_{\text{Mn}} \rangle = \langle N \rangle - \langle N_X \rangle$, as indicated by the horizontal double arrow in Fig. 3d. We then compute the relative fractions of intrinsic ($f_X$) and Mn-based ($f_{\text{Mn}}$) excitations using the expressions: $f_X = \langle N_X \rangle / \langle N \rangle$ and $f_{\text{Mn}} = 1 - f_X$.

In Fig. 3e,f, we present the dependence of the derived values of $\langle N_X \rangle$ and $\langle N_{\text{Mn}} \rangle$ on $\langle N \rangle$ (panel 'e'), along with the corresponding dependence of $f_X$ and $f_{\text{Mn}}$ on $\langle N \rangle$ (panel 'f'). At $\langle N \rangle < 1$, $\langle N_X \rangle$ is approximately equal to $\langle N_{\text{Mn}} \rangle$, and both increase linearly with $\langle N_X \rangle$. This indicates that the exciton has an equal probability of occupying either an intrinsic or Mn-base state ($f_X \approx f_{\text{Mn}} \approx 0.5$, Fig. 3f), suggesting that the rates of forward ($k_f$) and backward ($k_b$) transfer are also approximately equal in the single-exciton regime in the 'resonant' QD-Mn system[16,20].

This behavior changes once $\langle N \rangle$ exceeds 1. At this point, $\langle N_X \rangle$ begins to saturate, approaching a value of 2 as $\langle N \rangle$ increases to 21, while $\langle N \rangle$ continues to grow nearly linearly (Fig. 3e). A similar trend is observed in the plots of $f_X$ and $f_{\text{Mn}}$ versus $\langle N \rangle$ (Fig. 3f): while nearly identical at low excitation levels, the values of $f_X$ and $f_{\text{Mn}}$ begin to diverge sharply beyond $\langle N \rangle = 1$, indicating that an increasing fraction of photoinjected excitons are transferred to Mn ions.

These trends suggest that the ratio of forward to backward transfer rates ($k_f/k_b$) increases with $\langle N \rangle$, as more QDs becomes populated with biexcitons and higher order multiexcitons. As discussed in ref. [20], this behavior is expected for spin-conserving QD → Mn excitation transfer. Indeed, as shown in Fig. 2b, only one out of four spin configurations of a single exciton is suitable for transfer to the Mn ion (X → Mn* process). In contrast, the biexciton (XX) state always contains one suitable configuration (Fig. 3e, inset), implying that the forward transfer rate for XX → XMn* is four times higher than that for a single exciton. At the same time, the backward transfer rate (XMn* → XX) is reduced due to the lower probability of finding a spin-conserving configuration in the presence of a pre-existing exciton. Altogether, these considerations imply that the $k_f/k_b$ ratio increases with $\langle N \rangle$, leading to enhanced accumulation of excitations in Mn-based states.

**Scaling of SE-Auger Rate with Exciton Multiplicity**

Next, we examine the measured dependence of SE-Auger recombination rates ($k_{SE-A} = 1/\tau_{SE-A}$) on $\langle N \rangle$ (Fig. 4a, open diamonds). An initial hypothesis is to correlate $k_{SE-A}$ with the product of $\langle N_X \rangle \langle N_{Mn} \rangle$, since SE-Auger recombination involves interaction between intrinsic excitons and Mn-based excitations. However, applying this model results in a predicated dependence (dashed black line in Fig. 4a) that greatly overestimates the rate at which $k_{SE-A}$ increases with $\langle N \rangle$, compared to experimental data. This discrepancy arises from the nearly linear increase of $\langle N_{Mn} \rangle$ with $\langle N \rangle$, which drives a much faster growth in the theoretical $k_{SE-A}$ than is observed experimentally.

The observed discrepancy can be resolved by considering that SE-Auger recombination proceeds through a sequence of two correlated (ordered) spin-transfer steps (Fig. 4b). The first step involves a spin-down electron transfer from an excited Mn ion to a highly excited state in the QD. This is followed by a second step: the transfer of a spin-up electron from a QD band-edge state back to the Mn ion, filling a vacancy in its 3d shell.

The rate of the first step is enhanced by the high spectral density of available excited QD states and is further amplified when multiple excited Mn ions are present, acting as electron donors. In contrast, the second step is rate-limited due to the restricted number of suitable band-edge spin configurations and the involvement of only a single Mn acceptor ion – pre-selected during the first step (Fig. 4b). As a result, this second spin transfer becomes the rate-limiting step and ultimately determines the overall SE-Auger recombination rate.

These considerations lead to an important conclusion: the SE-Auger rate is independent of the number of excited Mn ions and is instead determined solely by the multiplicity of the intrinsic QD excitonic state. Based on the analysis of QD excitonic occupancies (Fig. 3e,f), the average exciton number $\langle N_X \rangle$ does not exceed 2. Therefore, the only multiexciton states that contribute to SE-Auger recombination are XMn* (hybrid biexciton) and XXMn* (hybrid triexciton).

Let the SE-Auger decay rates of these states be denoted as $k_{SE-A,1}$ and $k_{SE-A,2}$, respectively. Then, the total SE-Auger decay rate can be expressed as: $k_{SE-A,tot} = s_1 k_{SE-A,1} + s_2 k_{SE-A,2}$, where $s_1 = 2 - \langle N_X \rangle$ and $s_2 = \langle N_X \rangle - 1$ represent the probabilities that the QD is occupied by one or two intrinsic excitons, respectively. This analysis assumes the presence of at least one excited Mn ion and $\langle N_X \rangle$ lying between 1 to 2.

If we express $k_{SE-A,2}$ as $k_{SE-A,2} = \gamma k_{SE-A,1}$, the total Auger decay rate becomes $k_{SE-A,tot} = k_{SE-A,1} [\langle N_X \rangle (\gamma - 1) + 2 - \gamma]$. Here, $\gamma$ is a scaling factor accounting for the increased Auger rate with higher exciton multiplicity. Based on prior comparisons of exciton and biexciton SE energy transfer rates (Fig. 2b and Fig. 3e, inset), we estimate that $\gamma$ can be as large as ~4.

To analyze the experimental dependence of $\tau_{SE-A}$ vs. $\langle N \rangle$ (Fig. 4a, open diamonds), we apply this model using $k_{SE-A,1} = 0.95$ ps$^{-1}$, corresponding to a lifetime $\tau_{SE-A} = 1.05$ ps observed at the onset of SE-Auger decay. The time constant is modeled as: $\tau_{SE-A} = 1/k_{SE-A,tot}$. Fitting the data with $\gamma$ as an adjustable parameter yields good agreement between the model and experiment for $\gamma = 4.2$ (Fig. 4a, red line). As previously discussed, this value is close to the scaling factor for QD → Mn SE energy transfer. Based on this analysis, the SE-Auger lifetimes of hybrid multiexciton states XMn* and XXMn* are 1.05 ps and 250 fs (=1.05/4.2), respectively. Importantly, these constants are *not* affected by the presence of additional excited Mn ions.

The developed model also provides insight into the SE-Auger dynamics in wide bandgap QDs with a radius $r = 1.25$ nm, where the exciton energy $E_X$ (2.3 eV) exceeds the energy of a Mn spin-flip transition $E_{Mn}$ (2.1 eV), representing a *downhill* energy transfer regime (Fig. 5a, inset). Optical absorption and PL spectra of Mn-doped and undoped (reference) QDs of this size are shown in the bottom row of Fig. 2b, with their TA spectra presented in Supplementary Figs. 5 and 7.

Analysis of the early-time amplitude of the 1S bleach as a function of pump fluence for $\langle N \rangle$ values up to ~12 (Fig. 5a,b) reveals that the occupancy of intrinsic QD exciton states does not exceed one (Fig. 5c). This indicates that photoinjected excitations preferentially populate Mn-based states (Fig. 5d) – a behavior expected in the downhill regime[16], where forward QD → Mn energy transfer is favored over backward transfer due to a significant energetic driving force of $\Delta_{X-Mn} = 200$ meV (Fig. 5a, inset).

Therefore, from the onset of SE-Auger decay at $\langle N \rangle \approx 1$ up the maximum value of 12, the QD remains occupied by no more than one intrinsic exciton (Fig. 5c). Under these conditions, the SE-Auger lifetime is expected to remain constant, independent on pump fluence, and determined solely by $1/k_{SE-A,1}$. This prediction is supported by our lifetime measurements, which show a consistent, $\langle N \rangle$-independent SE-Auger lifetime of ~1.2 ps (Fig. 5e, main panel and inset). The corresponding decay rate, $k_{SE-A,1}$, = 0.83 ps$^{-1}$, is comparable to that obtained for the 'resonant' sample ($k_{SE-A,1}$ = 0.95 ps$^{-1}$). The apparent lack of dependence of the SE-Auger recombination rate of a hybrid XMn* biexciton on QD size likely reflects the short-range, contact-like nature of SE interactions.

**Coexistence of SE-Auger and Conventional Auger Recombination**

An intriguing case of coexisting conventional and SE-Auger recombination arises in a larger QD sample ($r$ = 2.25 nm), where the band-edge exciton energy ($E_X$ = 2 eV) lies below the Mn spin-flip transition energy ($E_{Mn}$ = 2.1 eV), resulting in a negative energetic driving force ($\Delta_{X-Mn}$ = –100 meV; Fig. 6a, right inset). This corresponds to an *uphill* regime that energetically favors intrinsic band-edge exciton states over Mn-based states. As a result, at the onset of the multiexciton regime ($\langle N \rangle \approx 1$), the QD depopulation dynamics are governed primarily by conventional Auger decay, characterized by $\tau_A$ = 230 ps (Fig. 6a, darker-colored traces). This time constant corresponds to a biexciton composed of two intrinsic 1S excitons ($X_{1S}X_{1S}$; Fig. 6a, right inset).

Interestingly, a comparison of long-time TA dynamics (1.5 ns time window) between Mn-doped and undoped samples shows no significant differences across all pump levels up to $\langle N \rangle$ = 65 (Fig. 6b), suggesting that conventional Auger recombination dominates in both cases. However, a closer look at the early-time dynamics reveals a sharp, short-lived spike in the doped sample at higher pump levels. This feature becomes pronounced when $\langle N \rangle$ exceeds ~3, as shown in the expanded early-time TA traces (Fig. 6a, lighter-colored traces), and is indicative of SE-Auger recombination.

At these high pump fluences, the $1S_e$ electron level becomes fully occupied, forcing excess photogenerated electrons into the higher-energy $1P_e$ state, located over 200 meV above the $1S_e$ level[28]. These 1P electrons, together with valence-band holes, form high-energy $X_{1P}$ excitons capable of exciting Mn ions via SE energy transfer as driving force becomes positive. This leads to the formation of a hybrid triexciton ($X_{1S}X_{1S}Mn^*$), which undergoes rapid SE-Auger recombination (Fig. 6a, left inset), resulting in the observed sub-picosecond TA decay component. Notably, the lifetime of this component (~200 fs) closely matches the SE-decay time of a hybrid triexciton in the 'resonant' sample (220 fs; Fig. 3b), again highlighting the lack of apparent dependence of SE-Auger lifetimes of QD size.

Coexistence of conventional and SE-Auger recombination is also observed in another 'uphill' sample ($r$ = 2.5 nm and $E_X$ = 1.98 eV). Over a 1.4 ns time window, the TA dynamics of doped and undoped QDs remain nearly identical for all pump levels up to ($\langle N \rangle$ = 140 Supplementary Fig. 8a). However, early-time traces (Supplementary Fig. 8b) clearly reveal sub-picosecond SE-Auger recombination, indicating the formation of hybrid multiexcitons via SE energy transfer from high-energy excitonic state.

Another distinctive feature of high-fluence TA dynamics in Mn-doped 'uphill' samples – absent in undoped references – is a pronounced *post-growth* of the 1S bleach signal following the SE-Auger spike (Fig. 6a, c and Supplementary Fig. 8b). This post-growth reflects repopulation of the $1S_e$ level by hot carriers generated during the SE-Auger process. These carriers subsequently undergo intraband relaxation (cooling) on a timescale of a few picoseconds (Fig. 6a), consistent with previously reported carrier cooling dynamics in CdSe QDs[29,30].

In addition to direct relaxation through the manifold of QD states, carrier cooling can also be mediated by Mn ions, which can temporarily trap a hot exciton and subsequently release it into a lower-energy QD state. Previous studies have shown that these processes occur on ultrafast timescales (~100 fs) and therefore do not introduce a significant delay into the intrinsic picosecond-scale intraband dynamics.

**Conclusions**

We have investigated multiexciton dynamics in a series of Mn-doped QD samples exhibiting distinct energetic alignments – *resonant*, *downhill*, and *uphill* – between the intrinsic band-edge exciton and the internal Mn spin-flip transition. We find that in both the *resonant* ($E_X \approx E_{Mn}$) and *downhill* ($E_X > E_{Mn}$) regimes, multiexciton dynamics are dominated by SE-Auger recombination. Increasing the pump level leads to a rapid (nearly linear) rise in the number of Mn-based excitations, while the occupancy of intrinsic excitonic states increases more moderately (sublinearly), saturating at about 1–2 excitons per QD. Under these conditions, the SE-Auger rate also shows only a modest increase, correlating with the number of intrinsic excitons per QD and showing no noticeable dependence on the number of excited Mn ions.

This behavior reflects the nature of the SE-Auger process, which involves two correlated (ordered) spin-transfer steps. The first is initiated by a spin-down electron transfer from a single (randomly selected) Mn ion, which must then also participate in the second step – a spin-up electron transfer that fills the vacancy in its 3d shell.

In the uphill regime, conventional Auger decay dominates at the onset of the multiexciton regime, when the average number of photoinjected excitons per dot ($\langle N \rangle$) is around 1 to 2. At higher pump levels, this process coexists with SE-Auger recombination, which appears as an ultrafast (sub-picosecond) decay component preceding the conventional Auger decay. In the uphill regime, we also resolve the repopulation of the band-edge $1S_e$ state due to intraband relaxation of a hot electron generated by the SE-Auger recombination.

The significance of the TA dynamics observed in 'uphill' samples lies in their ability to simultaneously reveal three distinct processes – conventional Auger recombination, SE-Auger recombination, and intraband relaxation – allowing for a direct comparison of their timescales.

This comparison shows that SE-Auger recombination is the fastest process, occurring on a sub-picosecond timescale. It proceeds two to three orders of magnitude faster than conventional Auger recombination (tens to hundreds of picoseconds) and approximately an order of magnitude faster than phonon-assisted cooling (a few picoseconds).

This latter finding is particularly notable, as phonon-mediated energy dissipation is typically regarded as the fastest and most unavoidable process in semiconductors. The ability of SE-Auger energy transfer to outpace this mechanism highlights its potential as a powerful tool for capturing hot-carrier kinetic energy that would otherwise be lost to phonons. This capability has already enabled demonstrations of high-efficiency SE-photoemission driven by visible light[20] and high-yield SE carrier multiplication (SE-CM)[31,32]. Looking forward, SE-Auger processes hold exciting promise for hot-electron photochemistry, particularly as drivers of demanding chemical reactions that require high reduction potentials.

**Acknowledgements**

This project was supported by the Solar Photochemistry Program of the Chemical Sciences, Biosciences and Geosciences Division, Office of Basic Energy Sciences, Office of Science, U.S. Department of Energy. V.P. acknowledges support by a LANL Director's Postdoctoral Fellowship.

**Author contributions**

V.I.K. conceived the idea and coordinated the overall research efforts. H.J. synthesized undoped and Mn-doped QDs. V.P. and C.L. performed the TA measurements. V.P. performed the MCD measurements. V.P. and V.I.K. analysed the data. V.P. and V.I.K wrote the manuscript with contributions from other co-authors.

**Competing interests**

The authors declare no competing interests.

# Figures

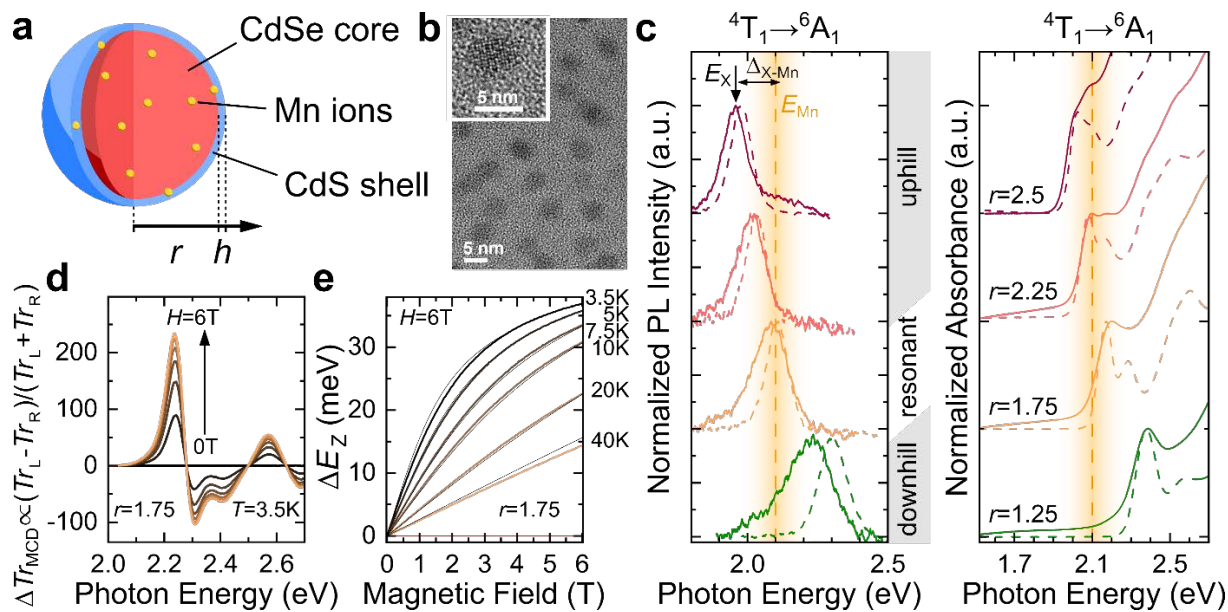

**Fig. 1 | Optical spectra and structural and magnetic properties of undoped and Mn-doped QDs. a,** Schematic representation of a QD, consisting of a size-controlled CdSe core (red sphere), doped with Mn ions (yellow dots), and enclosed within a thin protective CdS shell (blue layer). The CdSe core radius ($r$) ranges from 1.25 nm to 2.5 nm, and the CdS shell thickness ($h$) is approximately one semiconductor monolayer. **b,** Representative transmission electron microscopy (TEM) image of Mn-doped QDs with a core radius $r$ = 2.5 nm. Inset: an enlarge image on an individual QD revealing the underlying crystal lattice. **c,** Photoluminescence (left) and absorption (right) spectra of the samples studied in this work. Dashed and solid lines correspond to undoped and Mn-doped samples, respectively. The spectra of QDs of different sizes have been vertically shifted for clarity. The same color and style scheme for distinguishing between different samples is used throughout all figures in this work. A vertical orange stripe schematically represents a broadened Mn-ion spin-flip transition, with its center indicated by a dashed orange line. **d,** Representative magnetic field ($H$) dependence of magnetic circular dichroism (MCD) spectra of Mn-doped QDs with $r$ = 1.75 nm, measured at a temperature of $T$ = 3.5 with $H$ varied from 0 to 6 T in 1 T steps (see Methods for details of the MCD measurements). **e,** Zeeman splitting of the band-edge 1S QD transition, inferred from the measured MCD spectra, as a function of $H$, for sample temperatures from 3.5 to 40 K. Based on these data, the internal content of MCD-active Mn ions is determined to be 2%.

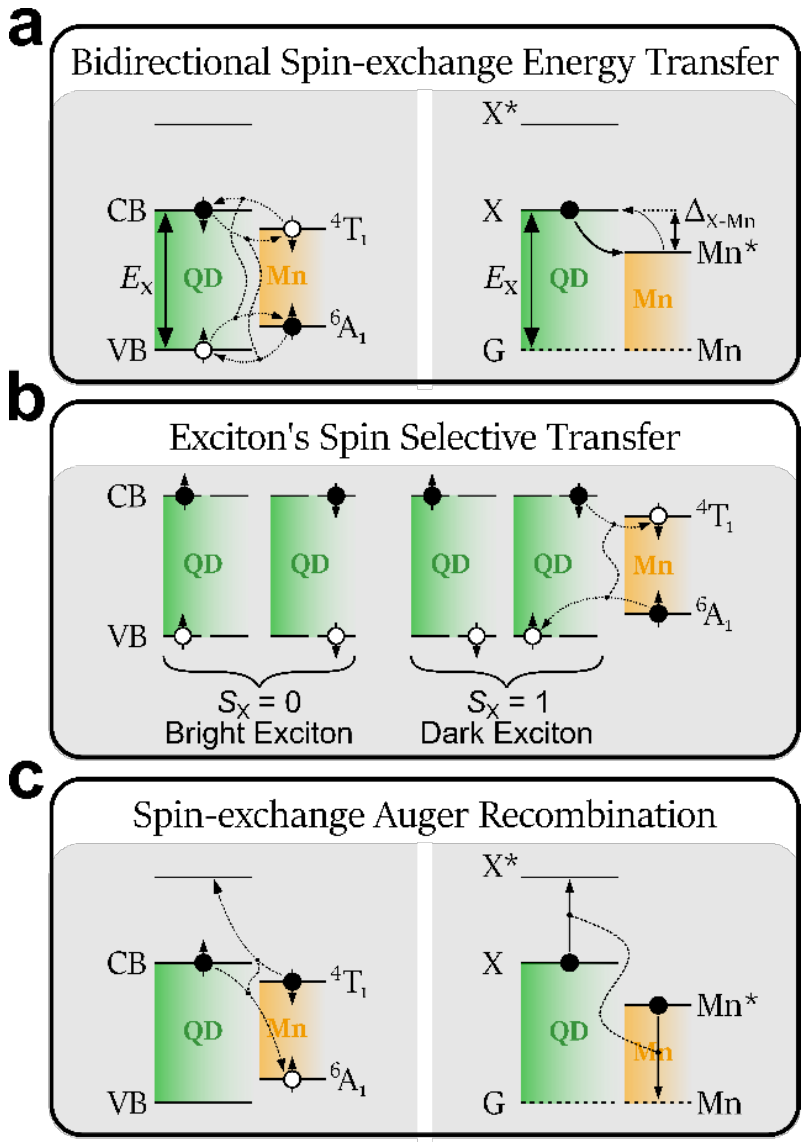

**Fig. 2 | Schematic illustration of spin-exchange (SE) -mediated processes in Mn-doped QDs. a,** Schematic of bidirectional SE excitation transfer, shown using spin-transfer (left) and exciton-transfer (right) representations. Forward (QD → Mn) and backward (Mn → QD) transfers are indicated by thick and thin arrows, respectively, in the right panel. **b,** A QD exciton state comprises four spin configurations: two corresponding to a bright (emissive) exciton with total spin $S = 0$, and two to a dark (nonemissive) exciton with $S = 1$. In our spin notation, the hole spin is defined by the spin of the electron that vacated the valence band to create the hole. Due to spin conservation, only one dark exciton state with the appropriate spin projection can participate in SE energy transfer with a Mn ion. As a result, the rate of QD → Mn SE energy transfer for an exciton state is one-quoter of that for a biexciton state, which always contains a spin configuration suitable for transfer. **c,** Schematic illustration of SE-Auger recombination, depicted using spin- (left) and exciton- (right) transfer representations.

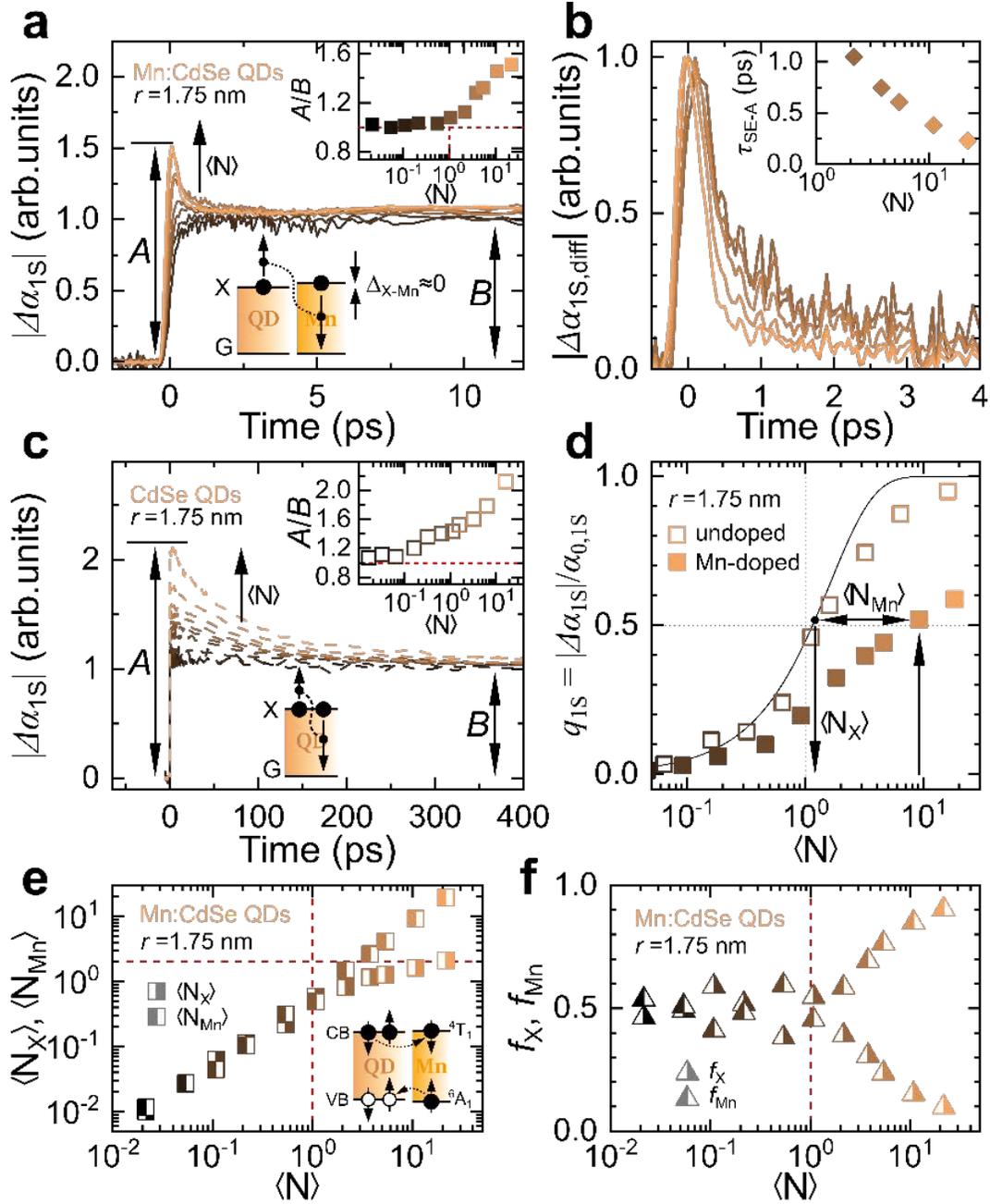

**Fig. 3 | Spin-exchange and conventional Auger recombination in the 'resonant' case. a,** Tail-normalized, pump-fluence-dependent 1S bleach dynamics obtained for Mn-doped 'resonant' QDs with $r$ = 1.75 nm, for which $E_X \approx E_{Mn}$ and hence $\Delta_{X\text{-}Mn} \approx 0$ (bottom inset: schematic of SE-Auger recombination). The data are shown as $|\Delta\alpha_{1S}|$ (pump-induced change in the 1S absorption coefficient) versus $\Delta t$ (pump-probe time delay). Measurements were performed at progressively increasing pump fluences, with $\langle N \rangle$ varied from ~0.02 to ~20 (higher $\langle N \rangle$ corresponds to lighter-colored TA traces). Top-right inset: Ratio of zero-delay ($A$) to long-delay ($B$) bleaching amplitude, plotted as a function of $\langle N \rangle$, indicating the onset of SE-Auger recombination at $\langle N \rangle \approx 1$. **b,** A TA component arising solely from SE-Auger recombination is obtained by subtracting the tail-

normalized single-exciton dynamics, recorded at $\langle N \rangle \ll 1$, from the TA traces measured for $\langle N \rangle >$ 1 (the resulting quantities are denoted $\Delta\alpha_{1S,\text{diff}}$). Inset: The time constant of SE-Auger recombination ($\tau_{\text{SE-A}}$) varies from ~1 ps to ~250 fs with increasing $\langle N \rangle$. **c,** TA measurements of Auger recombination in undoped 'resonant' QDs with $r = 1.75$ nm (reference sample). This panel uses the same representation of the TA data as for the Mn-doped sample in panel 'a' with one notable distinction – due to the much slower Auger recombination (~90 ps time constant), a much longer time window is used (400 ps versus 12 ps in panel '**a**'). Bottom inset: Schematic of conventional Auger recombination. Top-right inset: Plot of the *A/B* ratio versus $\langle N \rangle$ indicates the onset of conventional Auger decay at $\langle N \rangle \approx 1$, similar to SE-Auger recombination. **d,** $\langle N \rangle$-dependence of $q_{1S} = |\Delta\alpha_{1S}|/\alpha_{0,1S}$ ($\alpha_{0,1S}$ is the 1S ground-state absorption coefficient) for 'resonant' undoped (open squares; $q_{1S,\text{ref}}$) and Mn-doped (solid squares; $q_{1S,\text{Mn}}$) QDs with $r = 1.75$ nm. A solid line shows the dependence of $q_{1S}$ on $\langle N \rangle$ calculated assuming a Poisson distribution of QD excitonic occupancies. These measurements were used to determine how $\langle N \rangle$ is distributed between the intrinsic QD and Mn-based states ($\langle N_X \rangle$ and $\langle N_{\text{Mn}} \rangle$, respectively). Specifically, $\langle N_X \rangle$ is determined through three projections: $\langle N \rangle \rightarrow q_{1S,\text{Mn}} \rightarrow q_{1S,\text{ref}} \rightarrow \langle N_X \rangle$ (as indicated by arrows). Subsequently, $\langle N_{\text{Mn}} \rangle$ is found using $\langle N_{\text{Mn}} \rangle = \langle N \rangle - \langle N_X \rangle$. **e,** The dependence of $\langle N_X \rangle$ and $\langle N_{\text{Mn}} \rangle$ on $\langle N \rangle$ shows that while they are nearly identical at $\langle N \rangle < 1$, these quantities start to diverge at $\langle N \rangle > 1$. This divergence arises from an increasing fraction of QDs occupied with biexcitons, which exhibit a higher QD→Mn forward transfer rate compared to single excitons (inset). **f,** The corresponding dependences of the relative fractions of intrinsic and Mn-based excitations ($f_X$ and $f_{\text{Mn}}$, respectively) on $\langle N \rangle$ also show a divergence at $\langle N \rangle > 1$.

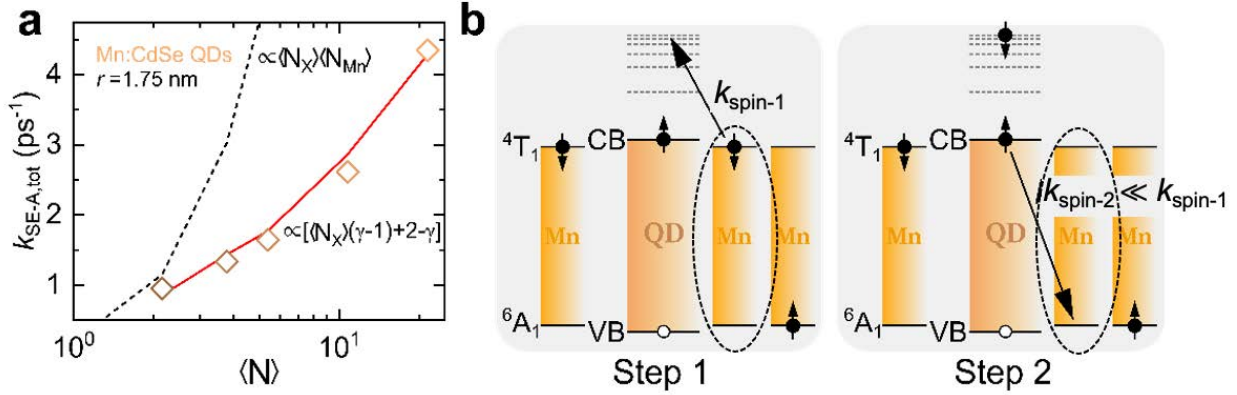

**Fig. 4 | Modelling of spin-exchange Auger recombination rates. a,** Open diamonds indicate SE Auger recombination rates ($k_{SE-A} = \tau_{SE-A}^{-1}$) determined from the measured time constants $\tau_{SE-A}$ for Mn-doped 'resonant' QDs (inset of Fig. 3b). The experimental trend is compared to modeling assuming $k_{SE-A}$ is proportional to the $\langle N_X \rangle \langle N_{Mn} \rangle$ product (dashed black line) or obtained using $k_{SE-A} = k_{SE-A,1}[\langle N_X \rangle(\gamma - 1) + 2 - \gamma]$ (red line). In this expression, $k_{SE-A,1}$=0.95 ps$^{-1}$ is the SE-Auger decay rate of a hybrid biexciton XMn* (determined from measurements), and $\gamma = k_{SE-A,2}/k_{SE-A,1}$ is a fitting parameter, where $k_{SE-A,2}$ is the SE-Auger decay rate of a hybrid triexciton (XXMn*). Based on the fit (red line), $\gamma = 4.2$, indicating $k_{SE-A,2} = 4$ ps$^{-1}$ and $\tau_{SE-A,2} = 250$ fs. **b,** Illustration of why the SE-Auger recombination rate depends solely on $\langle N_X \rangle$ and is not influenced by the number of excited Mn ions $\langle N_{Mn} \rangle$. The first step of SE-Auger recombination involves the transfer of a spin-down electron from one of the excited Mn ions to the QD. This step 'pre-selects' the ion for participation in the second spin transfer, as only this specific ion will have a vacancy in its 3d shell to accept a spin-up electron from the QD. The first step (rate $k_{spin-1}$) is faster than the second because it is facilitated by the large spectral density of high-energy QD states capable of accepting the electron from the Mn ion. Thus, the rate-limiting step is the second spin transfer (rate $k_{spin-2} \ll k_{spin-1}$), during which the electron moves from the QD to the pre-selected Mn ion. Since the first and second steps are closely coupled, $k_{spin-2} \approx k_{SE-A}$, where $k_{SE-A}$ is the overall SE-Auger recombination rate. As $k_{spin-2}$ depends solely on the occupancy of intrinsic QD states ($\langle N_X \rangle$) and not $\langle N_{Mn} \rangle$, the SE-Auger recombination rate $k_{SE-A}$ also depends only on $\langle N_X \rangle$.

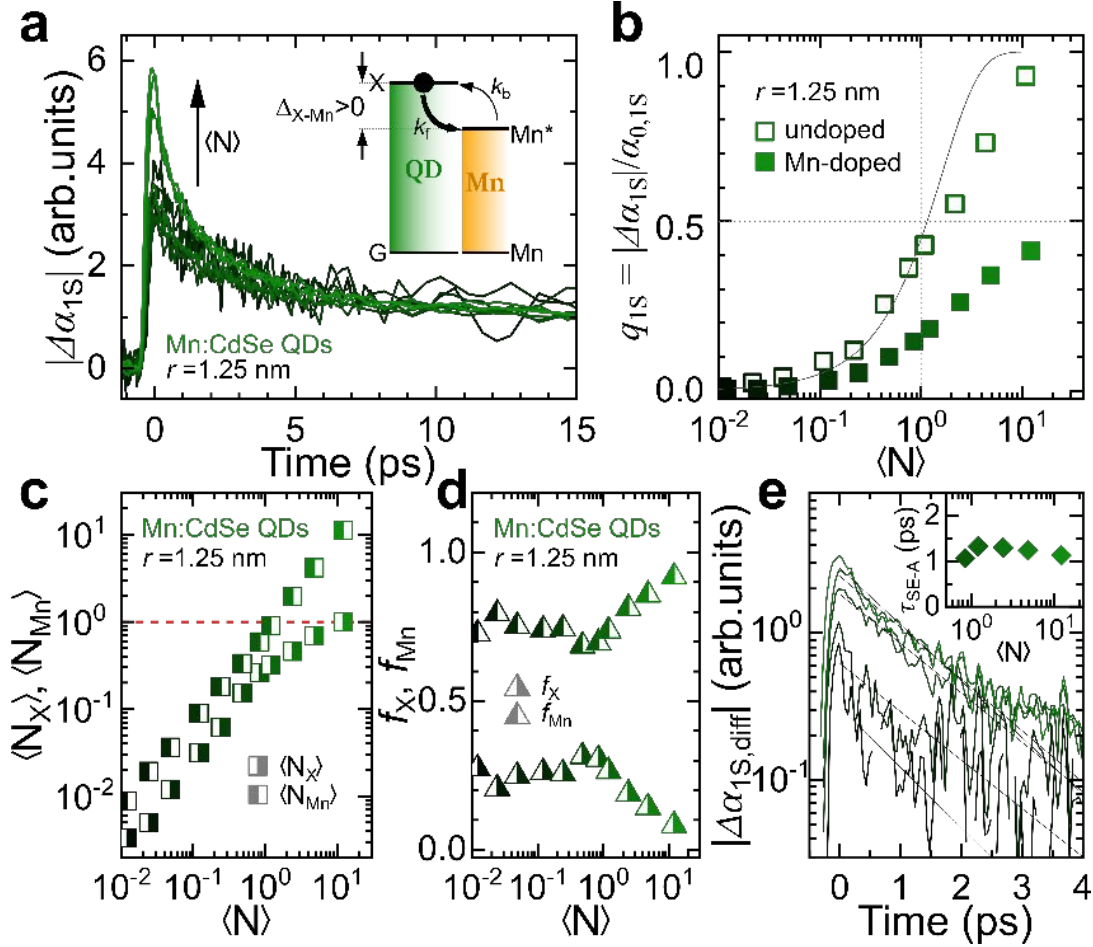

**Fig. 5 | SE-Auger recombination in the 'downhill' case. a,** Tail-normalized, pump-fluence-dependent 1S bleach dynamics of Mn-doped QDs with $r = 1.25$ nm, obtained for $\langle N_X \rangle$ varied from ~0.05 to ~12. For this sample, $E_X > E_{Mn}$ and hence $\Delta_{X\text{-}Mn} > 0$ (inset). This alignment of QD and Mn electronic levels favors forward QD → Mn excitation transfer over the backward Mn → QD transfer (indicated by thick and thin arrows, respectively, in the inset), reducing the occupancy of intrinsic QD states ($\langle N_X \rangle$) relative to Mn-based states ($\langle N_{Mn} \rangle$). **b,** This trend is apparent from measurements of the $\langle N \rangle$-dependence of $q_{1S} = |\Delta\alpha_{1S}|/\alpha_{0,1S}$, where $q_{1S}$ values for the Mn-doped sample (solids squares) are considerably lower than those for the reference undoped sample (open squares). The line corresponds to a theoretical $q_{1S}$-versus-$\langle N \rangle$ dependence assuming a Poisson distribution of QD excitonic occupancies. **c,** $\langle N \rangle$-dependences of $\langle N_X \rangle$ and $\langle N_{Mn} \rangle$, calculated based on the data in panel '**b**'. **d,** The corresponding $\langle N \rangle$-dependences of the relative fractions of excitations residing in intrinsic QD states ($f_X$) and Mn-based states ($f_{Mn}$). **e,** $\langle N \rangle$-dependent SE-Auger recombination dynamics, obtained from TA traces in panel '**a**' by subtracting tail-normalized dynamics measured at $\langle N \rangle \ll 1$. These dynamics are fitted to a single exponential decay (solid black lines), yielding a time constant of ~1 ps, which is virtually independent of $\langle N \rangle$ (inset).

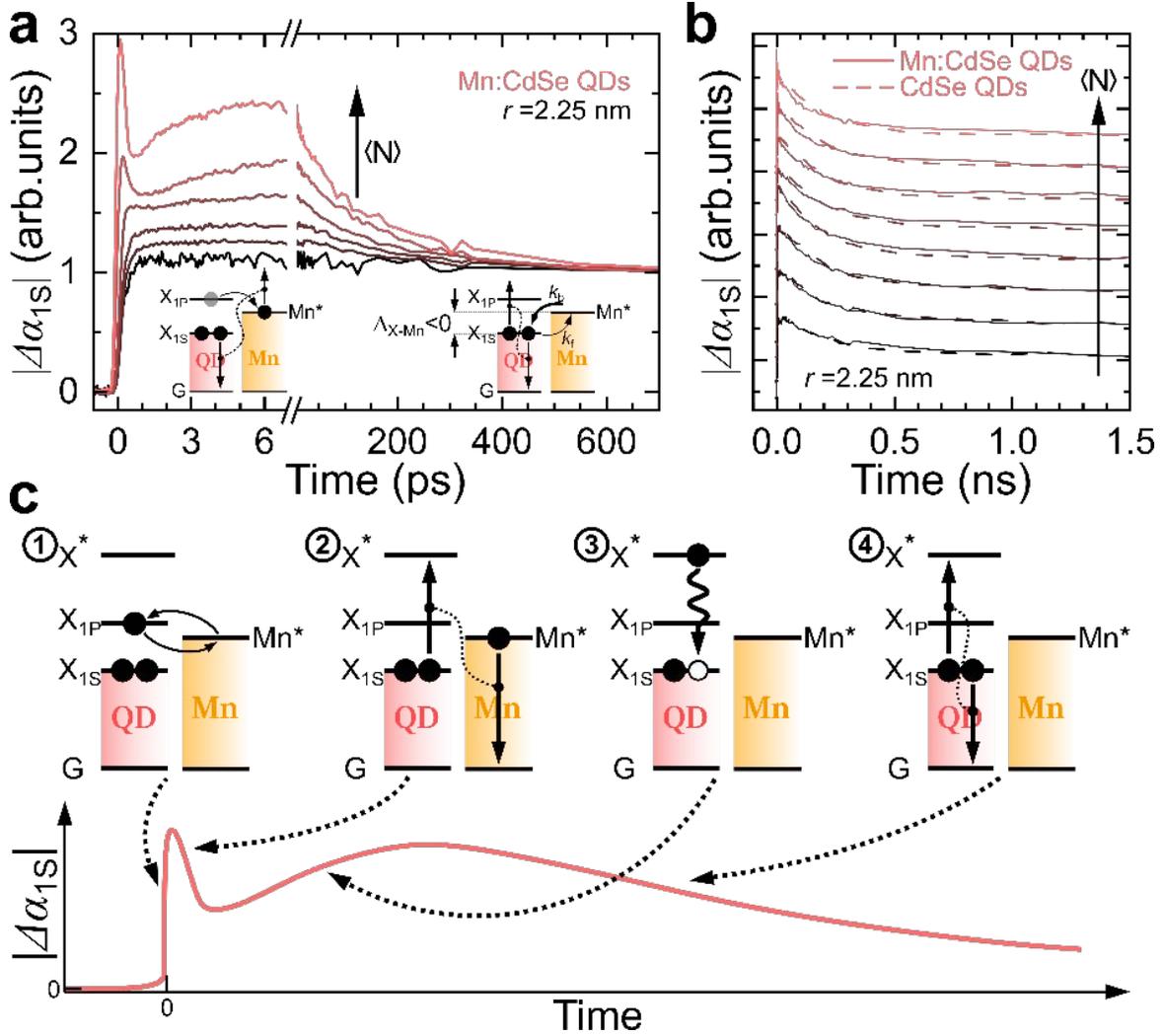

**Fig. 6 | Coexistence of conventional and SE-Auger recombination in the 'uphill' case. a,** Tail-normalized, pump-fluence-dependent 1S bleach dynamics of Mn-doped QDs with $r = 2.25$ nm, obtained for $\langle N_X \rangle$ varied from ~0.03 to 65 (note the time-axis break, separating an expanded view of short-time dynamics at $\Delta t < 7$ from the compressed view of long-time dynamics up to $\Delta t = 700$ ps). This sample corresponds to the 'uphill' configuration, where $E_X < E_{Mn}$ and $\Delta_{X-Mn} < 0$ (bottom-right inset). At low pump levels ($\langle N \rangle \ll 1$), TA traces are virtually flat across the entire time window ($\Delta t = 700$ ps), reflecting slow single-exciton decay. At intermediate pump levels ($\langle N \rangle$ of around 1 or slightly more), the TA traces exhibit an additional ~230-ps component (pronounced in the long-time dynamics after the time-axis break), characteristic of conventional Auger recombination. At these excitation levels, corresponding to the onset of the multiexciton regime, carriers are confined to the lowest-energy band-edge states, from which SE excitation transfer to Mn ions is not energetically feasible. Hence, multiexciton decay proceeds via regular Auger recombination (bottom-right inset). At $\langle N \rangle$ more than ~3, a very fast (sub-picosecond) component emerges in the TA traces (early time dynamics before the time-axis break), indicating the onset of SE-Auger recombination. The 1S bleach post-growth following the SE-Auger spike results from the repopulation of band-edge QD states due to relaxation of hot carriers produced by the SE-Auger recombination. **b,** Interestingly, when compared over a long time scale ($\Delta t = 1.5$ ns), pump-

intensity-dependent multiexciton dynamics of undoped and Mn-doped are virtually indistinguishable, both being governed by conventional Auger decay (in these measurements $\langle N_X \rangle$ varied from ~0.8 to 75). Only careful examination of near-time-zero dynamics of the Mn-doped sample reveals a sharp SE-Auger-decay-related spike, which is absent in the undoped sample. **c,** Schematic illustration of processes responsible for the various stages (1 to 4) of the 1S bleach temporal evolution for the 'uphill' Mn-doped sample shown in panels 'a' and 'b' (regime of high excitation levels, where ($\langle N \rangle > 3$). Stage 1: Formation of a hybrid triexciton ($X_{1S}X_{1S}Mn^*$) via $X_{1P}$ → Mn SE energy transfer. Stage 2: Sup-picosecond SE-Auger recombination of the hybrid $X_{1S}X_{1S}Mn^*$ state, generating a hot exciton ($X^*$) in the QD. Stage 3: Repopulation of the 1S band-edge states through relaxation of the hot exciton. Stage 4: Conventional Auger recombination of the $X_{1S}X_{1S}$ biexciton.

**Methods**

**Quantum Dot Synthesis**

*Chemicals.* We used chemicals as received without additional purification. Cadmium oxide (CdO, 99.99%), trioctylphosphine oxide (TOPO, 99%), trioctylphosphine (TOP, 97%), diphenylphosphine (DPP, 98%), 1-octadecene (ODE, 90%), oleic acid (90%), oleylamine (70%), sulfur powder (99.98%) and tributylphosphine (TBP, 95%) were purchased from Sigma-Aldrich. Octadecylphosphonic acid (ODPA, 99%) was purchased from PCI. Selenium (200 mesh, 99.999%) was purchased from Alfa Aesar. Manganese (II) acetate tetrahydrate ($Mn(ac)_2 \cdot 4H_2O$, 99.999%) was purchased from Strem.

*Synthesis of CdSe quantum dots (QDs)*. CdSe QDs with a wurtzite crystal structure were prepared following the protocol reported in ref. [22] without modifications.

Briefly, 60 mg of CdO, 280 mg of ODPA, and 3 g of TOPO were loaded into a 50 ml three-neck flask and degassed for 1 h at 150 °C. The mixture was then heated under inert an $N_2$ atmosphere to 320 °C and maintained at this temperature until it turned into a transparent solution, indicating full activation of the Cd precursors. Subsequently, 1.8 ml of TOP was injected, and the solution was further heated to 380 °C, at which point 0.433 ml of TOP-Se was added via a fast injection. After allowing the reaction to proceed for 1 min at 380 °C, it was quenched by cooling to room temperature.

QD size control was achieved performed by adjusting the amount of DPP added to the TOP used during the initial injection at 320 °C and the subsequent TOP-Se injection. The amounts of DPP used to prepare the 'downhill', 'resonant', and 'uphill' QDs were 0%, 1%, and 4%, respectively.

*Manganese incorporation.* To incorporate Mn ions into pre-synthesized CdSe QDs, we applied a diffusion doping procedure developed in ref. ref. [22] without modification.

Briefly, 5 ml of ODE, 0.433 ml of oleylamine, 0.20 ml of oleic acid, and 10 mg of $Mn(ac)_2 \cdot 4H_2O$ were loaded into a three-neck flask and degassed for 1 h at 120 °C. The mixture was then heated under an $N_2$ atmosphere to 320 °C, and 8.3 mg of CdSe QDs along with 0.667 ml of TBP-Se in ODE were swiftly injected. To reaction was allowed to proceed at 300 °C for 20–30 min, and was then quenched by cooling to room temperature.

*CdS shell deposition.* To enclose CdSe QDs (undoped and Mn-doped) within a protective CdS shell, we followed the procedure described in ref. [33] with slight modifications. Briefly, CdSe QDs were purified three time with methanol and redispersed in toluene.

The Cd precursor was prepared by degassing a mixture of 36.0 mg of CdO, 735 μl of oleic acid, and 3.24 ml of ODE at 100 °C for 1 h, followed by heating to 250 °C under an $N_2$ atmosphere and maintaining this temperature for an additional hour. The resulting colorless solution of Cd-oleate in ODE was then cooled to 100 °C.

The S precursor was prepared in a glove box by mixing 1.8 mL of TOP, 8.61 μL of DPP, and 4.43 mg of S powder, and stirring the mixture overnight.

Separately, 3.5 mL of ODE, 0.3 mL of oleylamine, and 0.15 mL of oleic acid were loaded into a three-neck flask and degassed at 100 °C for 1 hour. The purified QDs were then swiftly injected into the reaction mixture under an $N_2$ atmosphere. The temperature was raised to 220 °C, and 0.05 mL of Cd-oleate solution was added dropwise. After allowing the reaction to proceed for 2.5 minutes at 220 °C, 0.05 mL of the TOP-S solution was added dropwise, and the reaction was allowed to proceed for an additional 2.5 minutes. The injections of Cd and S precursors were repeated once more, after which the reaction was quenched by cooling to room temperature.

**Spectroscopic Characterization**

*Measurements of optical spectra.* For spectroscopic measurements, a QD solution was loaded into a 1 mm path length quartz cuvette.

Optical absorption spectra were acquired using a Perkin-Elmer Lambda 950 ultraviolet-visible-near infrared spectrophotometer. Solvent-related background was subtracted from the measured spectra.

Photoluminescence (PL) spectra were recorded using a Horiba Scientific FluoroMax+ spectrofluorimeter with excitation at 450 nm.

*Magnetic circular dichroism (MCD) measurements.* For MCD measurements, solid-state samples were prepared by incorporating QDs into a poly(lauryl methacrylate) matrix. The QD samples were placed into a variable-temperature insert (1.5–300 K) of a 7 T superconducting magnet with direct optical access (Oxford Spectromag). The samples were illuminated using a Xenon lamp, with the incident light mechanically chopped at 90 Hz. Light polarization was controlled using a Glan-Taylor polarizer (Thorlabs, GT15) and modulated at 50 kHz between right- and left-circular polarizations using a photoelastic modulator (PEM; Hinds Instruments, FS50). Measurements were performed with a magnetic field of up to 6 T applied along the probe light propagation direction (Faraday configuration) at various sample temperatures.

The transmitted light was spectrally dispersed by an HRS300 Teledyne spectrometer equipped with a 1200 groove-per-mm diffraction grating and detected using a silicon avalanche photodiode. The MCD signal ($\Delta Tr_{MCD}$) was obtained by measuring the ratio of the difference and sum of the transmitted light intensities for right and left circular polarizations ($Tr_R$ and $Tr_L$, respectively) using two separate lock-in amplifiers: one locked to the low-frequency mechanical chopper (sum signal) and the other to the PEM (difference signal). The MCD signal, calculated as $\Delta Tr_{MCD} \propto (Tr_L - Tr_R)/(Tr_L + Tr_R)$, is proportional to the Zeeman splitting ($\Delta E_Z$) between spin-dependent light-absorbing transitions.

***Transient absorption (TA) spectroscopy.*** Purified QDs dissolved in toluene were loaded into a 1 mm path length quartz cuvette. The sample concentration was adjusted to an optical density of approximately 0.15 at the 1S exciton peak position. TA measurements were performed using a standard pump–probe setup based on a regeneratively amplified ytterbium-doped potassium gadolinium tungstate (Yb:KGW) femtosecond laser (Pharos, Light Conversion) generating ~190 fs pulses at 1030 nm with a 500 Hz repetition rate.

The laser output was split into two arms: one seeding a second harmonics generator (HIRO HHG, Light Conversion) to produce 515 nm (2.41 eV) pump pulses, and the other directed toward a retroreflector mounted on a delay stage (maximum pump-probe delay, $\Delta t = 5$ ns) before exciting a YAG plate to generate broadband femtosecond white-light probe pulses. The pump pulse repetition rate was reduced to 250 Hz using a mechanical chopper with a 50% duty cycle, synchronized with laser pulses.

The pump and probe beams were focused and spatially overlapped on the sample, with spot diameters of ~120 μm and ~90 μm, respectively, as measured using a BladeCam DataRay beam profiler. The transmitted white light was detected by a fiber-coupled Avantes AvaSpec-Fast ULS1350F-USB2 spectrometer.

To prevent photocharging, the QD solutions were continuously stirred during the measurements using a stainless-steel rod driven by an external rotating magnet.

TA data were acquired by scanning the delay stage mounted in the probe arm. At each time delay, the difference between the transmission of white light under pump-on and pump-off conditions was recorded. The measured TA data were corrected for "chirp" (spectro-temporal broadening characteristic of femtosecond white-light pulses) as described in ref. x.

***Determination of absorption cross-sections and average QD excitonic occupancies.*** The average excitonic occupancy per QD ($\langle N \rangle$) generated by a pump pulse was determined using the relation $\langle N \rangle = \sigma_p(j_p/h\nu_p)$, where $\sigma_p$ is the sample-dependent absorption cross-section at the pump wavelength ($\lambda_p = 515$ nm), $j_p$ is the per-pulse pump fluence, and $h\nu_p = 2.41$ eV is the pump photon energy ($h$ is Plack's constant and $\nu_p$ the pump-photon frequency).

Absorption cross-sections were measured using QD samples before Mn doping, as in doped samples the occupancy of intrinsic QD states (measured via the TA method) is influenced by bidirectional transfer involving Mn-based states.

To quantify $\sigma_p$, we acquired a $j_p$-dependent late-time ($\Delta t = 3.5 - 4$ ns) 1S bleach signals ($\Delta\alpha_{1S}$) for $j_p$ varying from 1 μJ cm$^{-2}$ to 2.2 mJ cm$^{-2}$. At these time delays, all multiexcitonic species have already decayed via Auger recombination, so the measured TA signal is defined solely by the fraction of photoexcited QDs ($p_{exc}$) in the ensemble, independent of their initial excitonic occupancy.

Under Poisson photon absorption statistics[28], $p_{exc} = 1 - p_0 = 1 - \exp(-\langle N \rangle) = 1 - \exp(-\sigma_p \cdot j_p/h\nu_p)$, where $p_0$ is the fraction of unexcited QDs. The absorption cross-section $\sigma_p$ was extracted by fitting the measured $\Delta\alpha_{1S}$-versus-$j_p$ dependence with this expression.

**Data Availability**

Source data are provided for the main figures. The rest of the data is available from the corresponding author upon reasonable request.

**Methods-Only References**

# Supplementary Information for

# Interplay of Conventional and Spin-Exchange Auger Recombination in Magnetically Doped Quantum Dots

Valerio Pinchetti[1], Ho Jin [1,2], Clément Livache[1,3], and Victor I. Klimov[1]*

[1]Nanotechnology and Advanced Spectroscopy Team, C-PCS, Chemistry Division, Los Alamos National Laboratory, Los Alamos, New Mexico 87545, United States

[2]Department of Chemistry, Ulsan National Institute of Science and Technology, Ulsan, 44919, Republic of Korea

[3]Laboratoire de Physique de la Matière Condensée, Ecole Polytechnique, CNRS, IP Paris, 91128 Palaiseau, France

*Corresponding author's e-mail: klimov@lanl.gov

**Supplementary Notes**

**Supplementary Note 1: Analysis of magnetic circular dichroism data**

Magnetic circular dichroism (MCD) measurements were used to determine the magnetic field ($H$) -dependent Zeeman splitting ($\Delta E_Z$) of the band-edge 1S optical transition. For undoped QDs, $\Delta E_Z$ is related to $H$ by the following expression:

$$\Delta E_Z^{und} = g_{ex}\mu_B H, \qquad (S1)$$

where $g_{ex}$ is the exciton g-factor and $\mu_B$ is the Bohr magneton.

In Mn-doped QDs, the Mn ions act as magnetic dipoles that align under an external magnetic field, enhancing the total magnetic field experienced by the semiconductor host, thereby increasing the Zeeman splitting, In this case, it can be represented as:

$$\Delta E_Z^{Mn-doped} = g_{eff}\mu_B H + x_{Mn}N_0(\alpha - \beta)\langle S_J \rangle, \qquad (S2)$$

where $g_{eff}$ is the effective exciton g-factor, $N_0\alpha = 0.23$ eV and $N_0\beta = -1.27$ eV are the s-d and s-p exchange coupling constants for CdMnSe, respectively, and $\langle S_J \rangle$ is the Brillouin function with the argument is $g_{Mn}S_{Mn}\mu_B B(k_B T)^{-1}$. Here, $x_{Mn}$ is the relative content of MCD-active internal Mn ions, $S_{Mn} = 5/2$ is the Mn angular momentum, $g_{Mn}$ is the Mn g-factor, $k_B$ the Boltzmann constant, and $T$ is the sample temperature.

Equations (S1) and (S2) were used to analyze MCD data and, specifically, to determine the dopant contents and other magnetic characteristics of the undoped and Mn-doped QDs. The results of this analysis are summarized in Supplementary Table 2.

**Supplementary Figures**

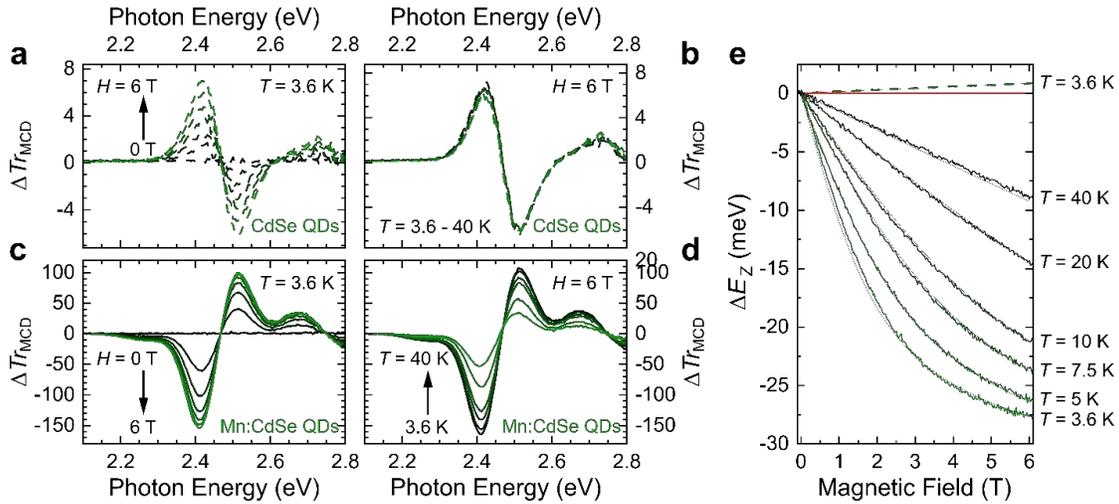

**Supplementary Figure 1. Magnetic circular dichroism (MCD) measurements on undoped and Mn-doped QDs with a CdSe core radius of 1.25 nm ('downhill' samples). a,** MCD spectra of undoped QDs measured at a fixed temperature of $T = 3.6$ K with the magnetic field varied from 0 to 6 T in 1 T steps. **b,** MCD spectra of undoped QDs measured at a fixed magnetic field of $H = 6$ T for varied sample temperatures ($T = 3.6, 5, 7.5, 10, 20,$ and 40 K). **c,** Same measurements as in panel '**a**' but for Mn-doped QDs. **d**, Same measurements as in panel '**b**' but for Mn-doped QDs. **e**, Magnetic field dependence of the Zeeman splitting of the band-edge 1S transition inferred from MCD measurements shown in panels '**a**'–'**d**' for undoped QDs (red line; $T = 3.6$ K) and Mn-doped QDs (green and black lines; $T$ varied from 3.6 K to 40 K). These data were used to determine the internal content of MCD-active Mn ions (Supplementary Note 1).

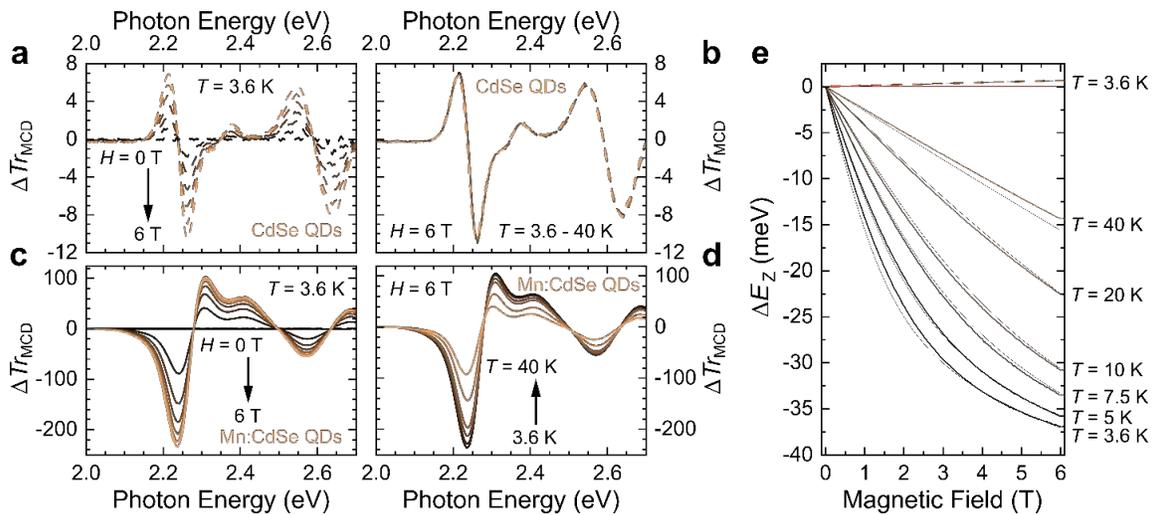

**Supplementary Figure 2. MCD measurements on undoped and Mn-doped QDs with a CdSe core radius of 1.75 nm ('resonant' samples).** This figure follows the same panel structure and presents the same types of information as Supplementary Fig. 1, but for QD samples with a different CdSe core radius ($r = 1.75$ nm).

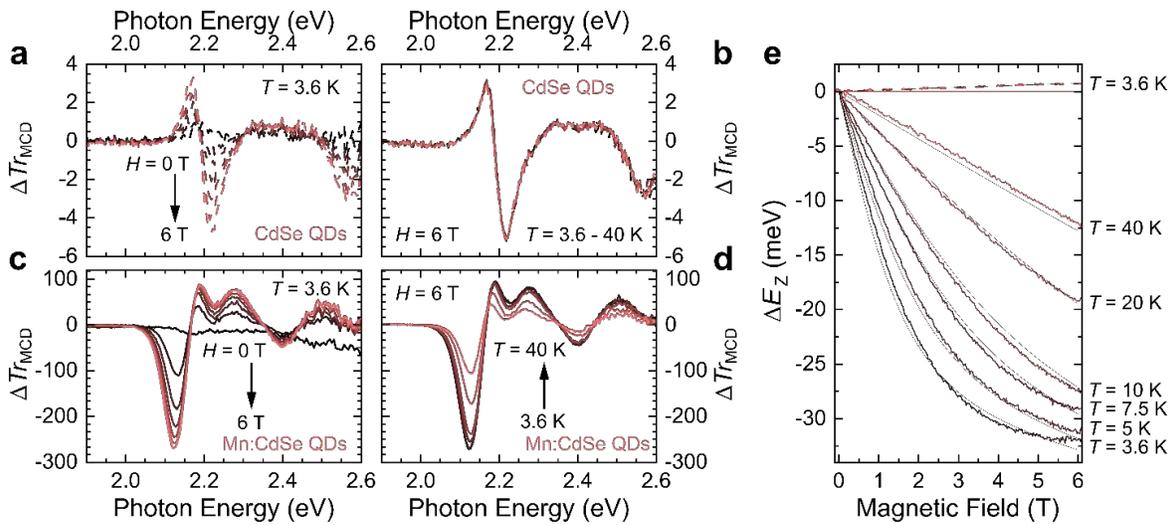

**Supplementary Figure 3. MCD measurements on undoped and Mn-doped QDs with a CdSe core radius of 2.25 nm ('uphill' samples).** This figure follows the same panel structure and presents the same types of information as Supplementary Fig. 1, but for QD samples with a different CdSe core radius ($r$ = 2.25 nm).

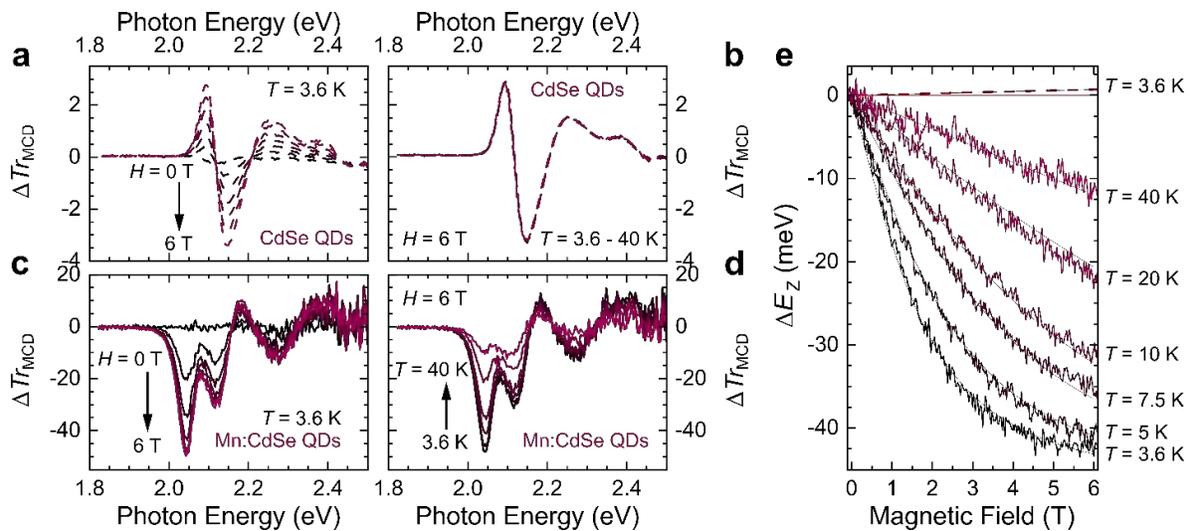

**Supplementary Figure 4. MCD measurements on undoped and Mn-doped QDs with a CdSe core radius of 2.5 nm ('uphill' samples).** This figure follows the same panel structure and presents the same types of information as Supplementary Fig. 1, but for QD samples with a different CdSe core radius ($r$ = 2.25 nm).

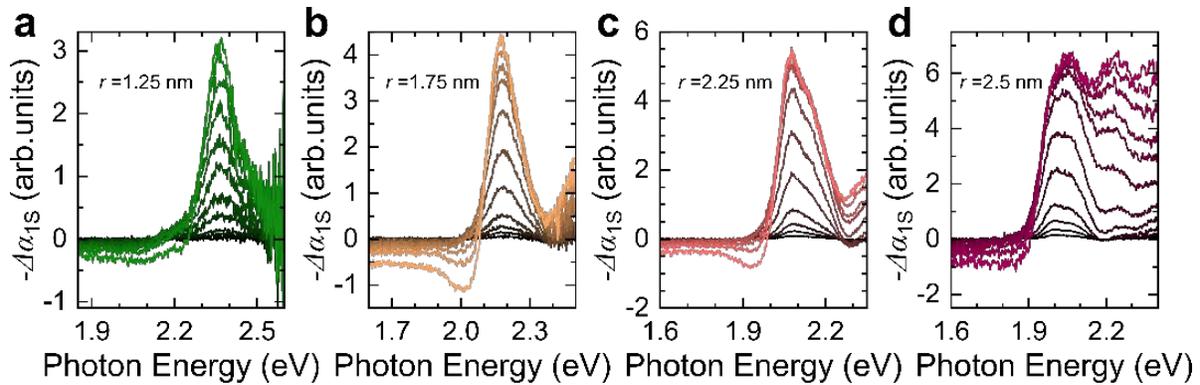

**Supplementary Figure 5. Pump-fluence-dependent transient absorption (TA) spectra of Mn-doped QDs with different CdSe core radii ($T$ = 300 K).** **a**, $r$ = 1.25 nm ('downhill' sample). **b,** $r$ = 1.75 nm ('resonant' sample). **c,** $r$ = 2.25 nm ('uphill' sample), and **d,** $r$ = 2.5 nm ('uphill' sample). The average QD excitonic occupancy ($\langle N \rangle$) varies from ~0.005 to ~12 (**a**), ~0.02 to ~50 (**b**), ~0.03 to 65 (**c**), and ~0.07 to ~180 (**c**).

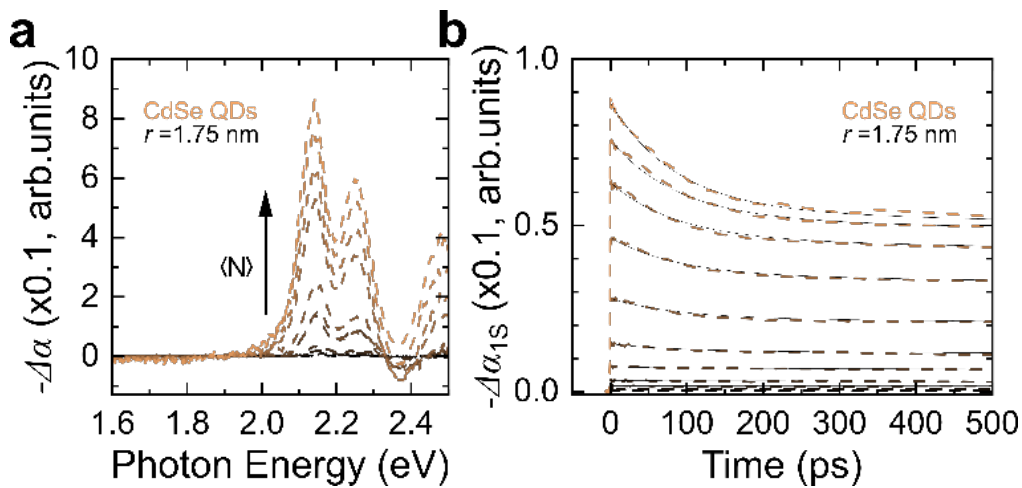

**Supplementary Figure 6. Transient absorption (TA) measurements of undoped QDs with $r$ = 1.75 nm ('resonant' sample)**. **a,** Pump-fluence-dependent TA spectra for $\langle N \rangle$ varied from ~0.01 to ~20. **b,** The corresponding 1S bleach dynamics (orange) are fitted with a double exponential decay (black), yielding a biexciton Auger lifetime of 88 ps.

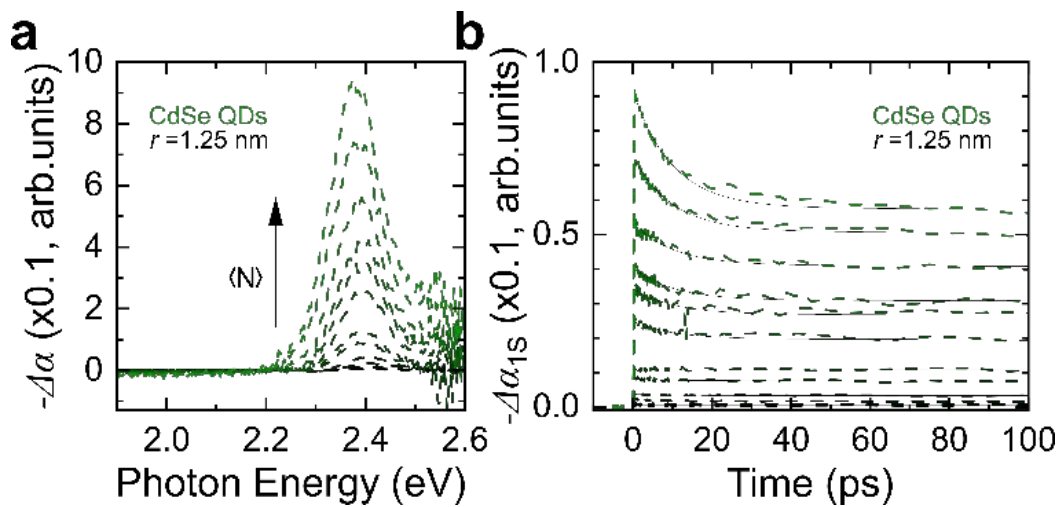

**Supplementary Figure 7. Transient absorption (TA) measurements of undoped QDs with $r$ = 1.25 nm ('downhill' sample).** This figure follows the same panel structure and presents the same types of information as Supplementary Fig. 6, but for a QD sample with a different CdSe core radius ($r$ = 1.25 nm). Fitting of the 1S bleach dynamics yields a biexciton Auger lifetime of 10 ps. $\langle N \rangle$ varies from ~0.004 to ~11.

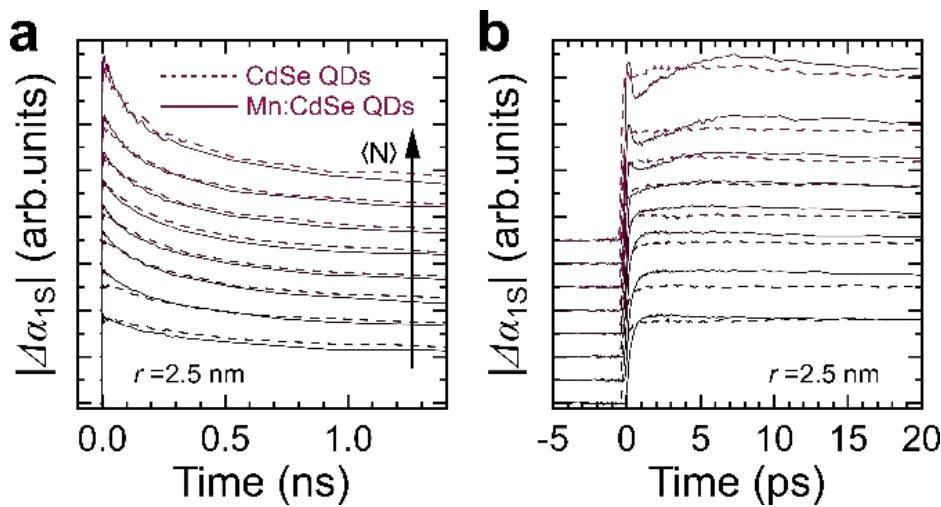

**Supplementary Figure 8. Comparison of TA dynamics of undoped and Mn-doped QDs with $r$ = 2.5 nm ('uphill' samples). a,** $\langle N \rangle$-dependent 1S bleach dynamics of undoped (dashed lines) and Mn-doped (solid lines) QDs shown over a 1.5-ns time window. **a,** Expanded view of the same dynamics over a 20-ps time window. $\langle N \rangle$ varies from ~1.5 to ~150.

**Supplementary Tables**

**Supplementary Table 1.** Summary of the spectroscopic and structural characteristics of undoped and Mn-doped QD samples studied in this work.

|  | 1S abs peak (eV) | PL peak (eV) | CdSe core radius (nm) | Internal Mn (%) | Internal Mn (number) |
|---|---|---|---|---|---|
| **CdSe QDs** | 2.378 | 2.299 | 1.25 | - | - |
| **Mn:CdSe QDs** | 2.389 | 2.230 | 1.25 | 1.7 | 1.2 |
| **CdSe QDs** | 2.172 | 2.104 | 1.75 | - | - |
| **Mn:CdSe QDs** | 2.203 | 2.086 | 1.75 | 2.05 | 3.9 |
| **CdSe QDs** | 2.082 | 2.035 | 2.25 | - | - |
| **Mn:CdSe QDs** | 2.098 | 2.014 | 2.25 | 1.9 | 7.8 |
| **CdSe QDs** | 2.026 | 1.977 | 2.5 | - | - |
| **Mn:CdSe QDs** | 2.007 | 1.957 | 2.5 | 0.3 | 1.6 |

**Supplementary Table 2.** Magnetic characteristics of undoped and Mn-doped QD samples studied in this work.

| | | **Undoped QDs** | **Mn-doped QDs** | | |
|---|---|---|---|---|---|
| | | $g_{ex}$ | $g_{eff}$ | $x_{Mn}$ | $g_{Mn}$ |
| **CdSe core radius** | 1.25 nm | 2.4 ± 0.1 | 9.3 ± 0.4 | 1.7 ± 0.1% | 2.05 ± 0.05 |
| | 1.75 nm | 2.0 ± 0.1 | 20.9 ± 0.7 | 2.0 ± 0.1% | 2.40 ± 0.05 |
| | 2.25 nm | 2.1 ± 0.1 | 13.6 ± 0.5 | 1.9 ± 0.1% | 2.60 ± 0.05 |
| | 2.5 nm | 2.0 ± 0.1 | 5.7 ± 0.4 | 2.9 ± 0.1% | 2.02 ± 0.05 |